\documentclass[3p,times]{elsarticle}

\usepackage{amssymb}

\usepackage{amsmath}

\usepackage[figuresright]{rotating}

\begin{document}

\begin{frontmatter}




\title{Enhancements of electron-positron pair production at very low transverse momentum in peripheral hadronic A + A  collisions}


\author[a]{M. C. Güçlü}
\author[b]{M. Yılmaz Şengül}

\address[a]{Department of Physics, Istanbul Technical University, Istanbul, Turkey}
\address[b]{Department of Electrical and Electronics Engineering, Haliç University, Istanbul, Turkey}

\begin{abstract}
The STAR collaboration has observed an excess production of electron-positron pairs which have transverse momenta $p_{\perp} <$ 150 MeV/c in peripheral gold-gold and uranium-uranium collisions.	
ALICE has also reported on an excess of $\mu^{+}\mu^{-} $ pairs at
low $p_{\perp}$ in very peripheral (70–90 \% centrality) lead-lead
collisions at a center-of-mass energy of 2.76 TeV/nucleon
pair. In literature, there are number of studies about the electromagnetic lepton pair productions in peripheral heavy-ion collisions. Almost all of them have a prediction that unitary is violated in these high energies so that probability of producing electron-positron pairs is greater than one. It is clear that at RHIC and LHC energies, probabilities of producing electron-positron pairs for impact parameters smaller than Compton-wavelength of electron is greater than one. This means that for small impact parameters where the electric field is strongest multiple electron-positron pair productions is inescapable and it must be included in calculations. In literature, there are number of explanations about the $p_{\perp}$  broadening however none of them has included the multi-pair production processes in their calculations. The aim of this paper is to show that electron-positron multi-pair production cross section is large and it can not be ignored in calculations.
\end{abstract}

\begin{keyword}


\end{keyword}

\end{frontmatter}


\section{Introduction}
\label{}

Dileptons play a crucial role in studying quark gluon plasma (QGP) created
at Relativistic Heavy-Ion Collider (RHIC) and at Large Hadron Collider (LHC). 
Although dileptons are produced in this hot and dense medium, their interactions
with this medium is very small. Therefore they carry direct information from
this QGP phase of the matter.

The ultraperipheral collisions (UPCs) of heavy ions are described
by an impact parameter b greater than the sum of the radius
of the colliding nuclei. If $ b= 0$ this is called head-on collision ($0\%$ centrality)
or if $b=2R_{A}$ it is called grazing collision ($100 \%$ centrality) where $R_{A}$
is the nuclear radius of the colliding nuclei.
When two ultrarelativistic heavy ions
pass near each other, the intense electromagnetic fields are
very strong so that they pull lepton pairs from the vacuum ($\gamma\gamma\rightarrow\ l^{+}l^{-}$ ).
Lepton pairs coming from the quark-gluon plasma can be
mixed with the electromagnetically produced lepton pairs.
Therefore, it is extremely important to investigate the electromagnetic
production of lepton pairs in detail. 

Electromagnetic lepton production
has been studied in UPCs by the STAR \cite{ja,a1}, ATLAS \cite{d}, ALICE \cite{a}, and CMS \cite{k} collaborations. The STAR collaboration has observed an excess production of electron-positron pairs which have transverse momenta $p_{\perp} <$ 150 MeV/c
in peripheral gold-gold and uranium-uranium collisions. The number of lepton pair production is more than the predictions from hadronic production models. 
The enhancement factors of the $p_{\perp}$ distributions of electron-positron pairs are the largest in 0.4-0.76 and 1.2-2.6 GeV/$c^{2}$ invariant mass regions for 60-80 $\%$ Au+Au and U+U collisions \cite{sy}. The observed excess is found to concentrate below $p_{\perp} \approx$ 0.15 GeV/c, on the other hand hadronic cocktail, are found for $p_{\perp} >$ 0.15 GeV/c in both mass regions of Au + Au
and U + U collisions. This excesses can not be explained by the QGP thermal radiation and $\rho $ in-medium broadening calculations. This is a sign of coherent photon–photon interactions in ultra-peripheral collisions \cite{zrty}. 

There are number of papers published to explain this excess of electron-positron pairs. S. R. Klein \cite{srk} calculated the cross sections and kinematics for
two-photon production of electron-positron pairs. The calculation is based on the STARlight simulation code, which is based on the Weizsäcker-Williams virtual photon approach. His results for the STAR continuum observations are compatible with two-photon production of $e^{+}e^{-}$
pairs. Klein at al. \cite{ks,kmxy,kmxy1}further derive an all order QED re-summation for the soft photon
radiation to investigate the electromagnetic production of lepton pairs with low total transverse momentum
in relativistic heavy-ion collisions. In this work, authors discussed the
additional $p_{\perp}$ broadening effects from multiple interactions with the medium
and the magnetic field contributions from the quark-gluon plasma for peripheral and central collisions.

In reference \cite{wpw,zbtx}, authors calculated the cross section of lepton pair production in the classical field approximation with equivalent photon approximation (EPA) as well as the corrections beyond EPA in the Born approximation. On the other hand, in ref. \cite{bzx,mlzz,wlpzw,ws}, authors discussed the advantages and inadequacies of EPA methods. Although EPA methods describe the $\gamma \gamma$ process relatively well, it has problems to describe various experimental results, such as the recently observed broadening of the pair transverse momentum with impact parameter. 

There are number of studies about the multiple lepton pair productions in the literature \cite{bht,kg,ahtb,baur,glu,gu} Almost all of them have a prediction that unitary is violated in these high energies so that probability of producing electron-positron pairs is greater than one. In Fig. 1,  it is clear that at RHIC and LHC energies, probabilities of producing electron-positron pairs for impact parameters smaller than Compton-wavelength of electron is greater than one. This means that for small impact parameters where the electric field is strongest multiple electron-positron pair productions is inescapable and it must be included in calculations. In literature, there are number of explanations about the $p_{\perp}$ broadening however none of them has included the multi-pair production processes in their calculations. The aim of this paper is to show that electron-positron multi-pair production cross section is large and it can not be ignored in calculations.

Ultra-relativistic peripheral collisions of heavy ions at RHIC and at LHC can produce
copies of numbers of lepton pairs via the two-photon process. Since the energies of the heavy ions are 200 GeV and 2760 GeV per nucleon in these colliders, multi-pair production cross sections of electron-positron pairs, are quite large so that it is possible to measure them experimentally. To calculate the multi-pair production probabilities, first, we should have an impact parameter dependence cross section. We have obtained a well-behaved impact parameter dependence cross section and by using the Monte Carlo methods, we have calculated multipair production cross sections
of electron-positron pairs in Au-Au and in Pb-Pb heavy-ion collisions at RHIC and at LHC energies, respectively. We have also used some experimental
restrictions in our calculation to compare our findings with the experimental results.

\section{Formalism}

Two heavy ions move with relativistic velocities $u$ along the $z-$axis in opposite directions to each other, and undergo the Lorentz contraction due to their velocities. The distance between the radii of the nuclei is written by the impact parameter $b$. Throughout this paper, we use the natural units with $\hbar=c=m=e=1$.

Since the physical processes such as lepton-pair production or annihilation occur with the interaction of fields, the lepton pair production from the electromagnetic field is expressed by the Interaction-lagrangian density depended on the classical $4$-vector potential $A^{\mu}$:
\begin{equation} \label{1}
	\mathcal{L}_{int}(x)=- \bar{\psi}(x) \gamma_{\mu} \psi(x) A^{\mu}(x)
\end{equation}
where $A^\mu$ can be written separately for two colliding nuclei $\mathcal{A}$ and $\mathcal{B}$, 
\begin{equation} \label{2}
	A^{\mu}(q,b)=A^{\mu}_{\mathcal{A}}(q,b)+A^{\mu}_{\mathcal{B}}(q,b)
\end{equation}

We can write the terms of 4-vector potential in momentum space as,

\begin{equation} \label{3a}
	A^0_{(\mathcal{A},\mathcal{B})}(q) =  -8 \pi^2  \gamma^2 (Ze)  G_E(q^2) f_z(q^2) \frac{\delta(q_0 \mp \beta q_z)}{q_z^2+\gamma_{\perp}^2} \exp \bigg(\pm i \mathbf{q_{\perp}} \cdot \frac{\mathbf{b}}{2} \bigg) , \\ 
\end{equation}
\begin{equation} \label{3b}
	A^z_{(\mathcal{A},\mathcal{B})}(q) = \pm \beta   
	A^0_{(\mathcal{A},\mathcal{B})}(q), \\ 
\end{equation}
\begin{equation} \label{3c}
	\mathbf{A}^{\perp}_{(\mathcal{A},\mathcal{B})}(q) = 0
\end{equation}

where $f_z(q^2)$ and $G_E(q^2)$ are the form factors. Form factors are crucial for the cross-section calculation of lepton-pair production in heavy-ion collisions. In our calculations, we have used the Fermi (Woods-Saxon) distribution for the charge distribution of the nucleus,
\begin{equation} \label{6}
	\rho(r)= \frac{\rho_0}{1+\exp \big( \frac{r-R}{a}\big)}
\end{equation}
where $R$ is the radius of the nucleus, and $a$ is a quantity related to the thickness of nucleus shell. Also, $\rho_0$ can be calculated by normalization. The Fourier transform of this charge distribution gives us the form factor of the nucleus
\begin{equation} \label{7}
	f(q^2)= \frac{4 \pi \hbar}{Zeq} \int r \rho (r) \sin(qr) dr
\end{equation}

Using the expressions given so far, the cross-section depending on the impact parameter $b$ for the lepton pair production can be written 

\begin{eqnarray}\label{15}
		\frac{d\sigma}{db} ={} &
		\frac{\pi}{8 \beta^2} \sum_{\sigma_k} \sum_{\sigma_q} \int_0^{\infty} qdq \, b J_0(qb) \,\int_0^{2 \pi} d \phi_q \int \frac{dk_z dq_z d^2 k_{\perp} d^2 K d^2 Q}{(2 \pi)^{10}}
		 \bigg\{ F \Big( \frac{\mathbf{Q}-  \mathbf{q}}{2}; \omega_{\mathcal{A}} \Big) F \big(-\mathbf{K};\omega_{\mathcal{B}} \big) \mathcal{T}_{kq} \Big(\mathbf{k}_{\perp}-\frac{\mathbf{Q}-\mathbf{q}}{2}; \beta \Big) \nonumber \\ &		 
		  + F \Big( \frac{\mathbf{Q}- \mathbf{q}}{2}; \omega_{\mathcal{A}} \Big) F \big(-\mathbf{K};\omega_{\mathcal{B}} \big) \mathcal{T}_{kq} \big(\mathbf{k}_{\perp}- \mathbf{K}; -\beta \big) \bigg\} \bigg\{F \bigg( \frac{\mathbf{Q} + \mathbf{q}}{2}; \omega_{\mathcal{A}} \bigg) F \big(-\mathbf{q}- \mathbf{K};\omega_{\mathcal{B}} \big) \mathcal{T}_{kq} \bigg(\mathbf{k}_{\perp}-\frac{\mathbf{Q}+ \mathbf{q}}{2}; \beta \bigg) \nonumber \\ & 		 
		  + F \Big(\frac{\mathbf{Q}+  \mathbf{q}}{2}; \omega_{\mathcal{A}} \Big) F \big(-\mathbf{q}- \mathbf{K};\omega_{\mathcal{B}} \big) \mathcal{T}_{kq} \big(\mathbf{k}_{\perp}+\mathbf{q}- \mathbf{K}; -\beta \big) \bigg\}
\end{eqnarray}
and this 10-dimensional integral can be calculated with Monte-Carlo integration method. Details of the calculations can be found in these articles \cite{kg,glu,gu}. 
Then, we can easily calculate the impact parameter $b$ dependence probability of producing lepton-pairs 
\begin{equation} \label{18}
	P(b)=\frac{1}{2 \pi b} \frac{d \sigma}{db} .
\end{equation}
However, for small impact parameters, the probabilities of lepton pair productions are greater than one, therefore unitary is violated. This means that there are more then one pair is produced and this process is calculated with the Poisson distribution to avoid the violation of unitary for small impact parameters
\begin{equation} \label{19}
	P_N(b)=\frac{P(b)^N e^{-P(b)}}{N!}
\end{equation}
where $N$ is the number of pairs. For one-pair cross section($N=1$), we just need to integrate the one-pair probability $P_1(b)$,
\begin{equation} \label{20}
	\sigma_{1_{pair}}=\int d^2 b \, P_1(b)
\end{equation}
In above derivations, we applied Feynman rules for the S-matrix. We first work with the operator form of the S-matrix and then take the matrix element with the initial (here the vacuum) and final (here N pairs) states. The Poisson form of the results
follows from the summation of a subset of all the diagrams that would occur in the external-field model \cite{glu}.

\section{Results}

We have calculated impact parameter dependence cross sections and found that for small impact parameters unitary is violated. This clearly can be seen in Fig. 1  so that for small impact parameters probabilities are greater than one. This means that more than one electron-positron pair is produced in single collisions of ions. Therefore we must consider this higher order corrections in our calculations. In Table I, total cross sections are shown for various collisions of ions at RHIC and LHC energies. Unitary violated cross sections are considerably greater than the one pair production cross sections and these unitary violated cross sections do not represent the real single pair cross sections. We can also clearly see that at these high energies and high charges of the colliding ions, two, three and even four pair productions probabilities are not negligible. It would be significant error without including these higher order effects in the calculations.

	\begin{table}[ht]
		\centering
		\caption{ By using Eq. 9, unitary violated total cross sections are obtained. With Poisson distribution multi-pair cross sections, total pair production cross sections results for different ions, energies are tabulated in unit of barn. It is important that unitary restored total pair cross sections are slightly smaller than the unitary violated cross sections and multi-pair cross sections are not negligible.}
		\label{tab:trigo}
		\begin{tabular}{|c|c|c|c|c|c|c|}
			\hline
			Ion&Unitary violated& N=1&N=2&N=3&N=4&Unitary restored total pair \\\hline
			RHIC Au+Au & 54400 & 51276& 1274 & 158 & 23 & 52672 \\\hline
			RHIC  U+U& 98300 & 90290 & 2964 & 508 & 109 & 93830 \\\hline
			LHC  Pb+Pb&226000 & 210700 & 5974 & 861 & 151 & 217493 \\\hline
		\end{tabular}
		\label{2}
	\end{table}

In Fig. 1 pair production probabilities of $N=1$, $N=2$ and $N=3$ pairs can be seen for $Au+Au$,$U+U$ and $Pb+Pb$ collisions for different energies, respectively. In each graph, first line represents the single pair production probabilities. These probability values are greater than $1$ that shows the contribution of more than one pair production probability can not be ignored especially for small impact parameters.
We have plotted the differential cross section as a function of transverse momentum of the produced pairs in Fig. 2. We have also calculated the integrated cross sections and ratio of the integrals between $0 - 0.15$ GeV/c and between 0 - infinity is nearly $1$

\begin{figure}
	\centering
		\centering
		\includegraphics[width=.5\linewidth]{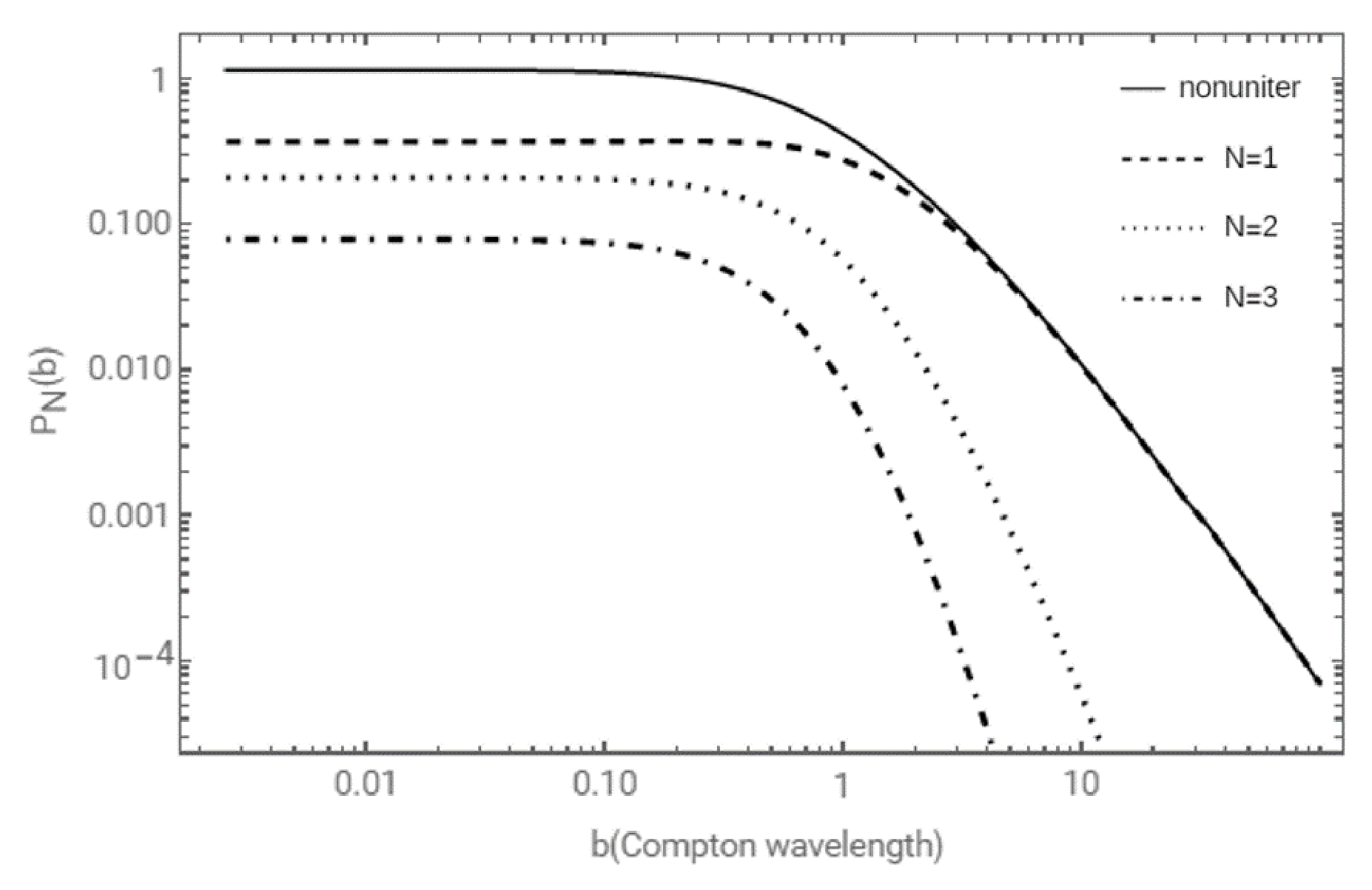}
		\centering
		\includegraphics[width=.5\linewidth]{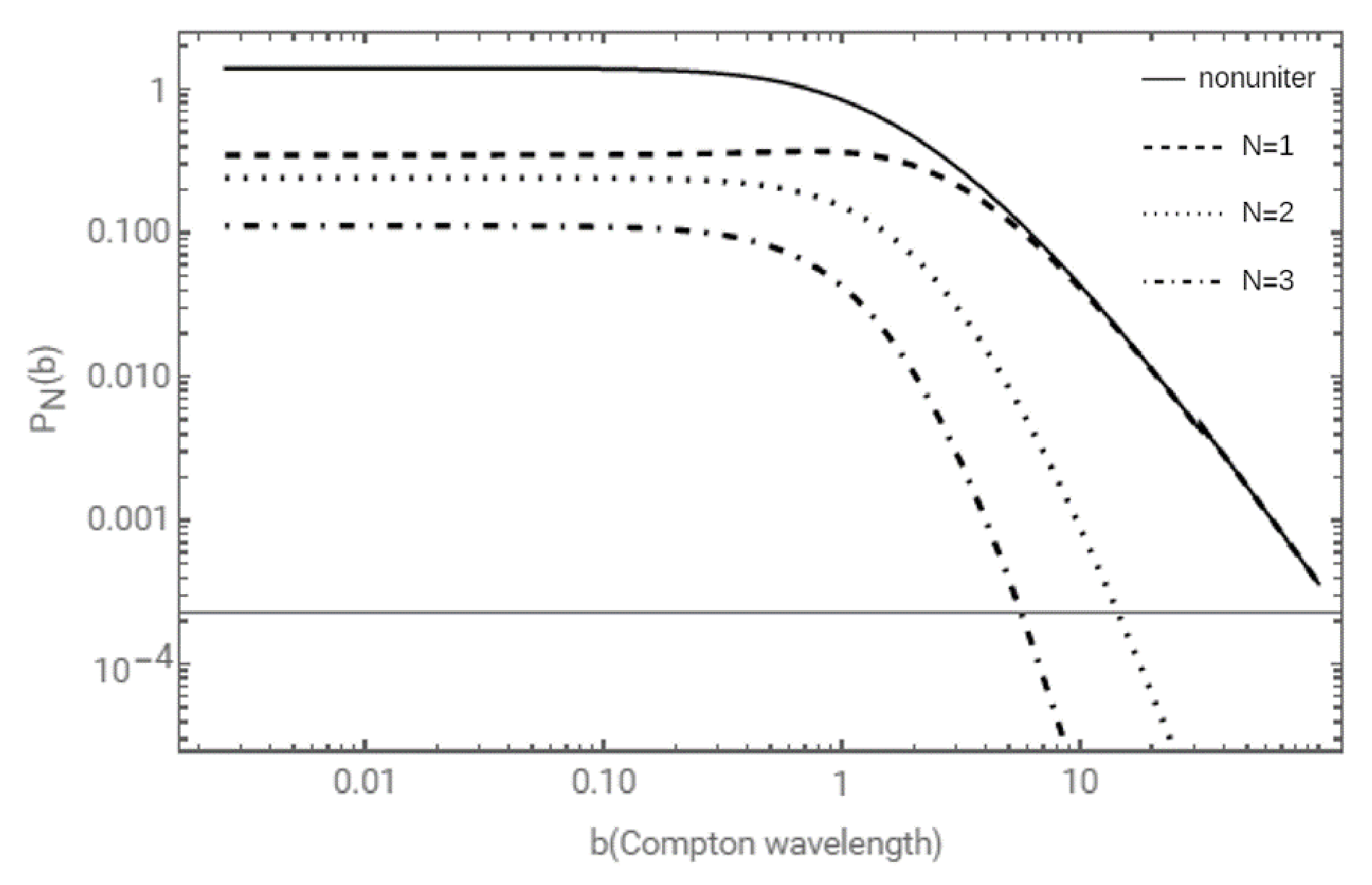}
		\centering
		\includegraphics[width=.5\linewidth]{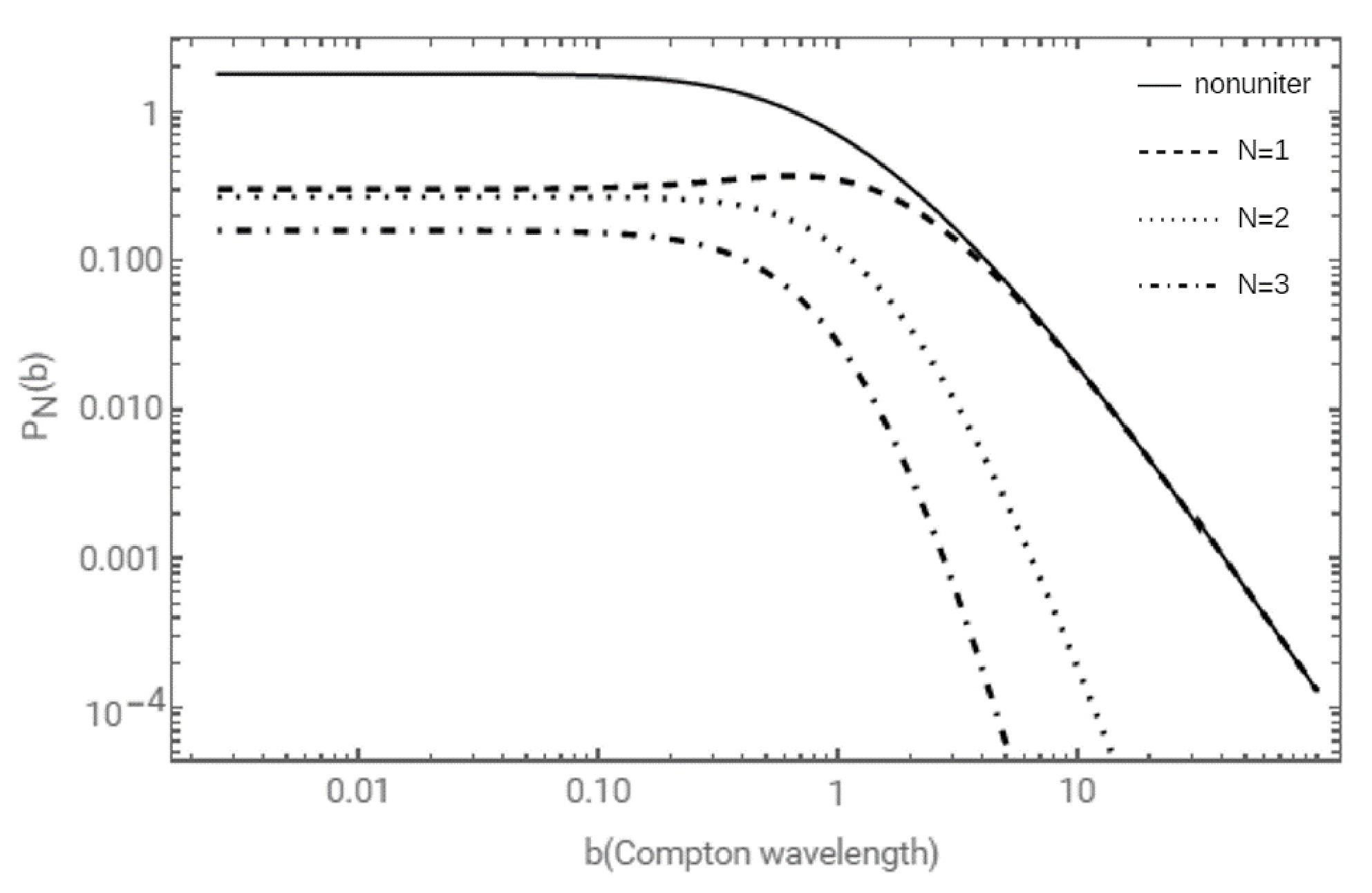}
	\caption{Probabilities of producing electron-positron pairs as a functions of impact parameters b (in unit of Compton wavelength of electron). Top graph is for the Au + Au collisions, middle one is the for U + U collisions and bottom one is for the Pb + Pb collisions. In all graphs, solid lines shows the nonuniter plots and it is clear that for small impact parameters they are greater than one. When we include the higher order effects, unitary is restored and for small impact parameters one-pair, two-pair and three-pair producing probabilities are significant.}
\end{figure}

\begin{equation} \label{ptratio}
	\frac{\int_{0}^{0.15 GeV/c} d p_{\perp} \frac{d\sigma}{d p_{\perp}}}{\int_{0}^{\infty} d p_{\perp} \frac{d\sigma}{d p_{\perp}}}\approx 1 .
\end{equation}
This means that transverse momentum of the produced leptons mostly in the range $0 - 0.15$ GeV/c. In fact, the maximum produced leptons transverse momentum are between $20-50$ MeV/c range. On the other hand, the ratio of the invariant mass $M_{e^{+}e^{-}}$ integrals between 0.4 $\leq M_{e^{+}e^{-}}\leq$ 2.6 $GeV/c^{2}$ and between 1.02 $MeV/c^{2}$ and $\infty$

\begin{equation} \label{mratio}
	\frac{\int_{0.4GeV/c^{2}}^{2.6 GeV/c^{2}} d M \frac{d\sigma}{d M}}{\int_{1.02 MeV/c^{2}}^{\infty} d M \frac{d\sigma}{d M}}\approx 0.24 - 0.27,
\end{equation}
where the ratio 0.24 is for Au + Au collisions at 200 GeV energies, and the ration 0.27 is for Pb + Pb collisions at 2.760 TeV energies. For the U + U collisions at 193 GeV energies this ration is about 0.245. Here 0.24 means that 24 $\%$ of the produced electron-positron pairs are in the range of invariant mass 0.4 $\leq M_{e^{+}e^{-}}\leq$ 2.6 $GeV/c^{2}$. This calculations clearly show that the ratios of different charges and energies are almost constant, however for the Pb + Pb collisions, the ratio is slightly higher. 

In Table II, we have calculated impact parameter range in terms of collision centrality. We have used a black-disk model \cite{srk} to convert from centrality to impact parameter range. We have done our calculations within the STAR acceptance: invariant mass of the pairs $M_{e^{+}e^{-}}>0.4$ GeV/$c^{2}$, pair rapidity $\mid y_{e^{+}e^{-}}\mid < 1$, individual lepton $p_{\perp,e} > 200 $ MeV/c and pseudo-rapidity $\mid \eta_{e} \mid < 1$. Our calculations show that unitary is violated for high energies and small impact parameters. For centralities smaller than the radius of the colliding nuclei, the unitary violation would be even greater. In Table II, on the right column we have tabulated the cross sections for certain impact parameter regions for various heavy ion collisions. In order to restore the unitary, we include the higher order corrections and calculated the one-pair, two-pair, three-pair and four-pair productions. It is clear that more than one-pair production cross sections are quite large and compatible with single-pair production. When we add all multiple pair productions, we can obtain the total pair production cross sections and these values are tabulated next to the far right column in the table.

\begin{figure}
	\centering
		\centering
		\includegraphics[width=.5\linewidth]{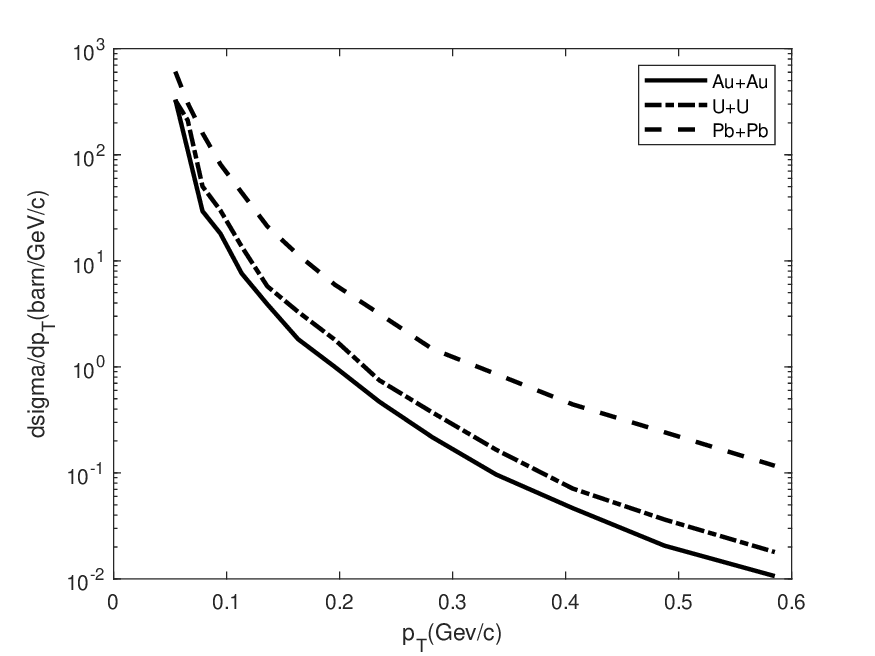}
		\centering
		\includegraphics[width=.5\linewidth]{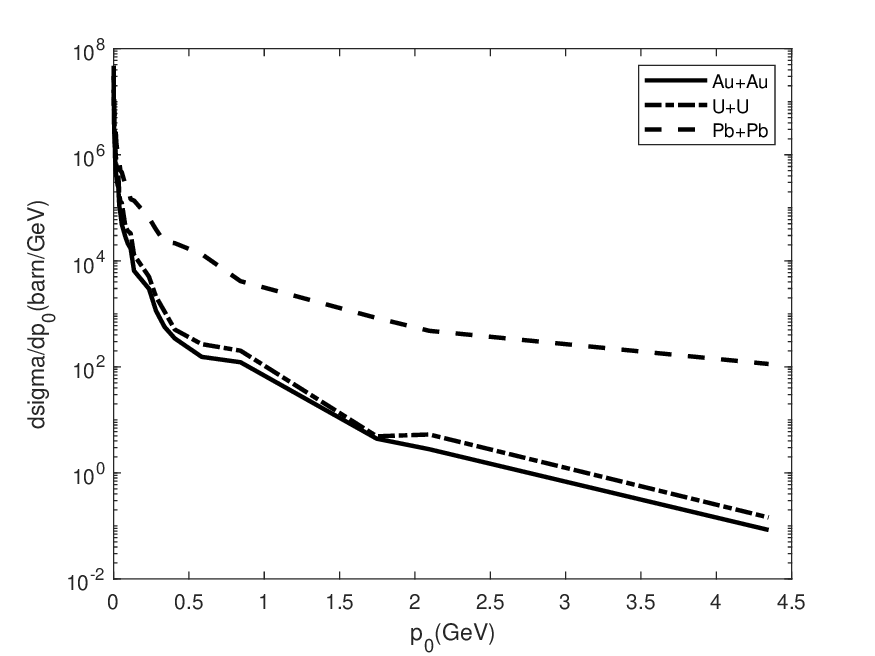}
		\centering
		\includegraphics[width=.5\linewidth]{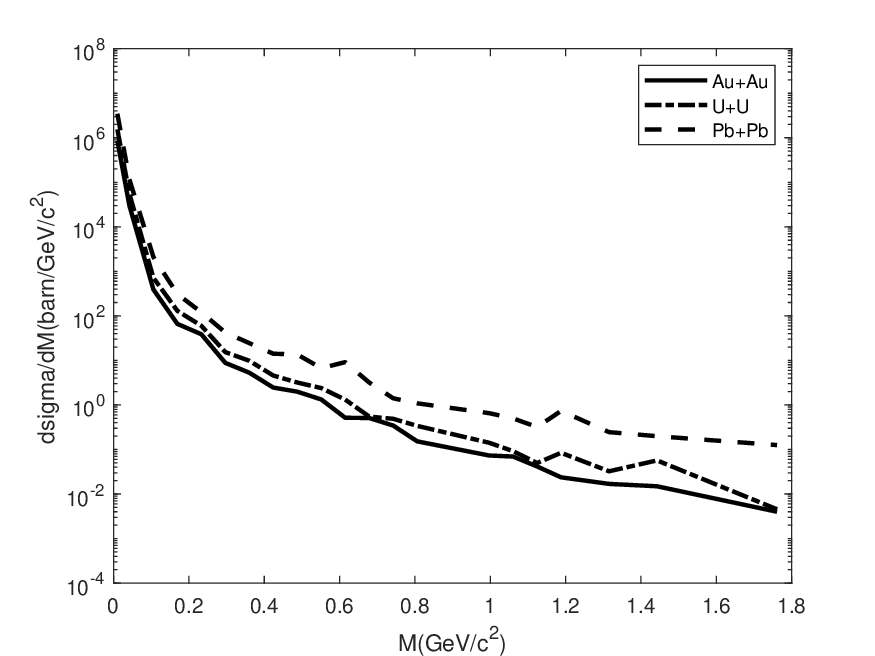}
	\caption{Differential cross sections as a functions of transverse momentum $p_{\perp}$, energy $p_{0}$ and invariant mass M of the produced electron-positron pairs. All three plots shows that when the colliding energies and charges of the heavy ions are large, the values of the differential cross sections are also large. In equations 12 and 13, the ratios of some certain integral ranges to the whole integral ranges are shown. Especially for the $p_{\perp} $ plot, it is interesting that almost all lepton pairs are produced below $p_{\perp} < $ 0.15 GeV/c. }
	\end{figure}

As expected, unitary violated cross sections are almost two times greater than total pair cross sections. Klein \cite{srk} has calculated the same calculations without including the higher order effects by using the STARlight Monte Carlo method. His calculations are 4 or 5 times larger than our total pair cross section calculations, 2 or 3 times greater than our unitary violated cross sections. One reason is that Weizsäcker-Williams method works well for impact parameter greater than one Compton wavelength of electron, not less than one Compton wavelength of electron. On the other hand, Monte Carlo QED calculations gives well behaved impact parameter dependence cross section, hence we can calculate the related observables directly without making further assumptions and approximations.   

\begin{table}[ht]
	\centering
	\caption{ Multipair cross sections, total pair production cross sections and unitary violated cross sections results for different ions, energies and impact parameter ranges.}
	\label{tab:trigo}
	\begin{tabular}{|c|c|c|c|c|c|c|c|}
		\hline
		Ion&b range&N=1(mb)&N=2(mb)&N=3(mb)&N=4(mb)&Total pair(mb)&Unitary violated(mb) \\\hline
		RHIC Au+Au & 11.4-13.2 fm & 0.508 & 0.287 & 0.108 & 0.031 & 0.942& 1.57 \\\hline
		RHIC Au+Au & 9.4-11.6 fm  & 0.530 & 0.300 & 0.113 & 0.032 & 0.983& 1.64 \\\hline
		RHIC Au+Au & 4.8-9.4 fm   & 0.749 & 0.424 & 0.160 & 0.045 & 1.39 & 2.33 \\\hline
		RHIC  U+U  & 14.1-15.8 fm & 0.481 & 0.426 & 0.252 & 0.111 & 1.33 & 2.83 \\\hline
		LHC  Pb+Pb & 13-14.7 fm   & 0.513 & 0.356 & 0.164 & 0.057 & 1.11 & 2.05 \\\hline
	\end{tabular}
	\label{1}
\end{table}

\section{Conclusion}

We have calculated impact parameter dependence cross sections of lepton pair production from electromagnetic fields of relativistic heavy ions. At these energies and small impact parameters it is clear that unitary is violated. This means that, second order perturbation theory is not sufficient alone to describe the lepton pair production. When we include the higher order effects in the calculations, table 1 and 2 shows that multiple pair production cross sections are not negligible.

In table 1, we can see that unrestricted multi-pair cross sections makes about $\sim$ 2 or 3 percent of the total pair cross sections, however for the restricted case in table 2, multi-pair production cross sections makes about more than half of total pair cross sections. For the case of U + U collision, one pair production cross section is almost equal to the two pair production cross section, since the charge of the uranium is larger than gold nuclei and cross section is proportional to  $\sim Z^{4}$. We can also explain why multi-pair cross sections are compatible with the single pair production cross sections at 70 - 90 $\%$ centrality collisions because at this impact parameters electric fields of the heavy ions reaches the maximum values.  

We have also calculated differential cross sections as a function of transverse momentum of produced lepton pairs. Electron-positron pairs are created mostly in the transverse momentum domain of less than  $0.15$ MeV/c and maximum probability of the pairs are produced between $20 - 50$ MeV/c transverse momentum region. When we include multi-pairs in our calculations this can explain the excess production of electron-positron pairs which have transverse momenta $p_{\perp} <$ 150 MeV/c in peripheral Au + Au and U + U collisions.  

In reference \cite{grss} the authors study the invariant-mass distributions of dileptons produced in ultrarelativistic heavy-ion collisions at very low pair transverse momenta, $p_{\perp} <$ 150 MeV/c. 
In their calculations, they show that the combination of photon fusion, thermal radiation and final-state hadron decays gives a fair description of the low-$p_{\perp}$ invariant-mass as well as $p_{\perp}$ spectra as recently measured by the STAR collaboration for different centrality classes. The coherent contribution dominates in peripheral collisions, while thermal radiation shows stronger increase with centrality. 

Recent study \cite{lzz,a2,s} about the impact parameter dependence of the cos4$\phi$
azimuthal asymmetry for purely electromagnetic lepton
pair production in heavy ion collisions at low $p_{\perp}$ shows that
azimuthal asymmetry has a strong $b_{\perp}$ dependence, specifically, the asymmetry decreases with increasing impact parameter. 

As a final remark, there was an experimental attempt \cite{vane} to measure double-pair production. Electron-positron pairs created by
fully stripped 200 A GeV sulfur ions incident on thin targets of Al, Pd, and Au. Since the colliding energies of heavy ions are not as large as nowadays, multiple pairs were not seen, however upper limits on the total cross section were set. We have produced a new Monte Carlo technique for calculating the impact
parameter-dependence of the cross section which also allows a calculation of the multi-pair production cross section. We find our results are consistent with the upper limits which experiments \cite{vane} put on this process. With the advent of higher-energy heavy-ion colliders, the study of the physics of
the two-photon process emerges as an exciting new field. The process may be used as a means of production of exotic particles perhaps the study of nonperturbative effects in QED. The ability to calculate the impact parameter dependence of the cross section and the related ability to calculate multiple-pair production provides an additional piece in the understanding of the process.






\begin{thebibliography}{00}
\bibitem{ja}
	J. Adams et.al.,
	Phys. Rev. C,
	70 (2004) 031902(R).
\bibitem{a1}
	J. Adam et.al.,
	Phys. Rev. Lett.,
	121 (2018) 132301.
\bibitem{d}
    M. Dyndal (ATLAS Collaboration),
	Nucl. Phys. A
	967 (2017) 281.
\bibitem{a}
	E. Abbas et al. (ALICE Collaboration), 
	Eur. Phys. J. C,
	73 (2013) 2617.
\bibitem{k}
	V. Khachatryan et al. (CMS Collaboration), 
	Phys. Lett. B,
	772 (2017) 489.
\bibitem{sy}
    S. Yang,
    Int. Jour. of Mod. Phys.: Conf. Ser.,
    46 (2018) 1860013.	
\bibitem{zrty}
    W. Zha, L. Ruan, Z. Tang, Z. Xu, S. Yang,
    Phys. Lett. B,
    781 (2018) 182.
\bibitem{srk}
	S. R. Klein,
	Phys. Rev. C,
	97 (2018) 054903.
\bibitem{ks}
	S. R. Klein, P. Steinberg,
	Ann. Rev. of Nucl. and Part. Sci.,
	70 (2020) 323.
\bibitem{kmxy}
    S. Klein, A. H. Mueller, B. Xiao, F. Yuan,
    Phys. Rev. Lett.,
    122 (2019) 132301.
\bibitem{kmxy1}
    S. Klein, A. H. Mueller, B. Xiao, F. Yuan, 
    Phys. Rev. D,
    102 (2020) 094013.
\bibitem{wpw}
	R. Wang, S. Pu, Q. Wang,
	Phys. Rev. D,
	104 (2021) 056011.
\bibitem{zbtx}
	W. Zha, J. D. Brandenburg, Z. Tang, Z. Xu,
	Phys. Lett. B,
	800 (2020) 135089.
\bibitem{bzx}
	J. D. Brandenburg, W. Zha, Z. Xu,
	The Eur. Phys. Jour. A,
	57 (2021) 299.
\bibitem{mlzz}
	Z. Ma, Z. Lu, J. Zhu, L. Zhang, 
	Phys. Rev. D,
	104 (2021) 074023. 
\bibitem{wlpzw}
	R. Wang, S. Lin, S. Pu, Y. Zhang, Q. Wang,
	Phys. Rev. D,
	106 (2022) 034025.
\bibitem{ws}
	W. Schäfer,
	The Eur. Phys. Jour. A,
	56 (2020) 231.
\bibitem{bht}
    G.Baur, K. Hencken, D. Trautmann
    Physics Report 453 (2007) 1.
\bibitem{kg}
	S. Karadağ, M. C. Güçlü,
	Phys. Rev. C,
	102 (2020) 014904. 
\bibitem{ahtb}
	A. Alscher, K. Hencken, D. Trautmann, G. Baur,
	Phys. Rev. A,
	55 (1997) 396.
\bibitem{baur}
    Phys. Rev. A
    42 (1990) 5736.
\bibitem{glu}
	M. C. Güçlü, J. Li, A. S. Umar, D. J. Ernst, and M. R. Strayer,
	Ann. Phys. (NY) 272, 7 (1999).
\bibitem{gu}
	M. C. Güçlü, 
	Nuc. Phys. A 668, 149 (2000).
\bibitem{grss}
    M. Kłusek-Gawenda, R. Rapp, W. Schäfer, A. Szczurek,
    Phys. Lett. B,
    790 (2019) 339.
\bibitem{lzz}
    C. Li, J. Zhou, Y. Zhou,
    Phys. Rev. D,
    101 (2020) 034015.
\bibitem{a2}
    J. Adam et.al.,
    Phys. Rev. Lett.,
    127 (2021) 052302.
\bibitem{s}
    A. M. Sirunyan et al.,
    Phys. Rev. Lett.,
    127 (2021) 122001.
\bibitem{vane}
    C. R. Vane, S. Datz, E. F. Deveney, P. F. Dittner, H. F. Krause, R. Schuch, H. Gao, R. Hutton,
    Phys. Rev. A, 56 (1997) 5.


\end{thebibliography}

\section{Acknowledgements}
This research is partially supported by Istanbul Technical University.	

\end{document}